\documentclass[11pt]{article}
\usepackage{amsmath}
\usepackage{amsfonts}
\usepackage{dsfont}
\usepackage{multirow}
\usepackage{amssymb}

\def\We{e}
\def\Ge{\mathfrak{e}}

\newtheorem{theorem}{Theorem}

\def\be{\begin{equation}}
\def\ee{\end{equation}}
\def\bea{\begin{eqnarray}}
\def\eea{\end{eqnarray}}

\begin{document}

\thispagestyle{empty}


\ 
\vspace{0.5cm}

\begin{center}

{\Large{\sc{Classification of real three-dimensional Poisson-Lie groups}} }

\end{center}

\medskip

\begin{center} \'Angel Ballesteros, Alfonso Blasco, Fabio Musso
\end{center}

\begin{center} {\it {Departamento de F\'{i}sica,  Universidad de Burgos, 
09001 Burgos, Spain}}

e-mail: angelb@ubu.es, ablasco@ubu.es, fmusso@ubu.es
\end{center}

  \medskip

\begin{abstract} 
\noindent
All real three dimensional Poisson-Lie groups are explicitly constructed and fully classified under group automorphisms by making use of their one-to-one correspondence with the complete classification of real three-dimensional Lie bialgebras
given in~\cite{gomez}. Many of these 3D Poisson-Lie groups are non-coboundary structures, whose Poisson brackets are given here for the first time. Casimir functions for all three-dimensional PL groups are given, and some features of several PL structures are commented.
 \end{abstract}

\bigskip\bigskip\bigskip\bigskip

\noindent
PACS: 02.20.Qs \quad 02.20.Uw \quad 02.30.Ik

\noindent
KEYWORDS: 
\newline Poisson-Lie groups, Lie bialgebras, Hopf algebras, integrable systems, Lie algebras, Lie groups, Poisson algebras, Casimir functions, quantum deformations


\section{Introduction}

Poisson-Lie (PL) groups are Poisson structures on Lie groups for which the group multiplication is a Poisson map, and play an outstanding role in the theory of classical integrable models (see~\cite{Drinfeld}-\cite{YKS} and references therein). Indeed, they were initially 
introduced by  Drinfel'd~\cite{Drinfeld} to give a geometrical description of the Poisson algebra defined by the elements of
the transition matrix for a large class of Hamiltonian systems that are integrable through the inverse scattering method.
Later on, Semenov-Tian-Shansky~\cite{STS} established the connection between PL groups and the group
of dressing trasformations of a completely integrable system (see also \cite{YKS}). Moreover, 
quantum groups are just Hopf algebra quantizations of PL groups~\cite{DrICM}-\cite{Bidegain}, quantum spaces are the quantizations of their corresponding Poisson homogeneous spaces~\cite{STS,zakr2, poishom,BHOSpl}, and the Poisson analogues of quantum algebras can be understood as PL structures defined on the dual group~\cite{Marmo,BBMbook}. It is also worth to stress the relevance of certain PL groups and their Lie bialgebras in the theory of T-dual sigma models~\cite{Klimcik}-\cite{Marian} and in (2+1) gravity~\cite{AMII}-\cite{BHMplb}. Moreover, being instances of Poisson-Hopf algebras, PL groups can be used to construct $N$-particle integrable systems through the so-called coalgebra symmetry method (see \cite{confseries}-\cite{Io2}) and some specific PL structures on solvable groups have been recently shown to arise as the Poisson manifolds underlying the integrability of a large class of Lotka-Volterra systems~\cite{LV}.

As established by Drinfel'd, there exists a one to one correspondence between the Poisson-Lie structures on a (connected and simply connected) Lie group $G$ and the Lie bialgebra structures $(g,\delta)$ on its Lie algebra $g$. Therefore, the classification problem for PL structures on $G$ is the same as the classification (under automorphisms) of the Lie bialgebra structures on $g=\mbox{Lie}(G)$. Complementarily, we recall that if a given Lie bialgebra $(g,\delta)$ is a coboundary one, this means that the cocommutator $\delta$ is obtained from a classical $r$-matrix, and in this case the full PL group structure associated to $(g,\delta(r))$ is obtained through the Sklyanin bracket given by $r$. For simple Lie algebras, this
problem has been studied in~\cite{BelDr,Stolinb}; in
this case, all Lie bialgebras are of the coboundary type and
their classification reduces to obtain all 
constant solutions of the classical Yang--Baxter equation. 
However, for non-semisimple groups non-coboundary Lie bialgebras can exist and for them no Sklyanin bracket is available. In this case, the PL group associated to a given non-coboundary Lie bialgebra has to be obtained by solving the compatibility conditions between the group product and the Poisson bracket and by imposing that the linearization of the latter corresponds to the Lie bialgebra under consideration (see~\cite{euclideo,galileo,anna,brihaye,checos}).

The aim of this paper is to complete the classification and the explicit construction of PL structures on all real 3D Lie groups by taking into account that the full classification of real 3D  Lie bialgebra structures was performed in the remarkable paper~\cite{gomez}. It turns out that the vast majority of Lie bialgebras found in~\cite{gomez} for the non-semisimple cases were non-coboundaries and, to the best of our knowlege, the PL groups associated to many of these non coboundaries have not been explicitly constructed so far.
Throughout the paper, we will follow the notation from~\cite{pavel,Mubar}, that classifies all real 3D Lie algebras into nine classes called $A_{3,i}$ ($i=1,\dots,9$), that we will appropriately connect with the classification used in~\cite{gomez} (see~\cite{Snobl} for the translation into the original Bianchi classification). 

Therefore, in this work we present the new classifications of PL groups corresponding to the solvable Lie algebras $A_{3,2}$, $A_{3,4}$ (the (1+1) Poincar\'e algebra), $A_{3,5}$ and $A_{3,7}$, together with the corresponding Casimir function for each of the PL brackets.
We recall that all the 3D real coboundary PL groups have been explicitly constructed in~\cite{rezaei} through their corresponding $r$-matrices, and all these results will be also recovered here in a more general framework. Also, PL structures on the Heisenberg-Galilei(1+1) group $A_{3,1}$ were obtained in~\cite{Kuper} and their classification in terms of ther corresponding Lie bialgebra structures was performed in~\cite{BHPheis, galileo}; the classification of PL structures on the Euclidean group $A_{3,6}$ was given in~\cite{euclideo} and the `book group' $A_{3,3}$ structures have been recently analysed in~\cite{BBMbook} through the same metodology that we will follow in the present paper and with the emphasis put on their applications in integrable PL dynamics. We emphasize that all these known results will be recovered from a common computational perspective and by adding the Casimir functions for all the PL structures. 

The paper is organized as follows. In section \ref{methods} we tersely recall the relevant theory of Poisson-Lie groups needed
for the scope of this paper and, simultaneously, we also describe the methodology that we have used in order to get the results that we will present in the sequel. In section \ref{results} we give
the complete classification for Poisson-Lie groups corresponding to each of the unequivalent real 3D Lie algebras. For each of these nine Lie groups we find by direct computation the most generic (multiparametic) Poisson bracket that is compatible with the group multiplication. Afterwards, we will compare the (dual of the) linearization of such generic PL bracket with the classification of Lie bialgebra structures on the corresponding Lie algebra given in~\cite{gomez}, thus obtaining the equivalence classes of PL structures. Among them, both the coboundary and non-coboundary cases are identified, and the Casimir functions for all of them are computed. In order to illustrate the method, 
the new classification for the 3D solvable group generated by the Lie algebra $A_{3,2}$ is firstly explained in detail, and for the rest of the cases we present the results in a schematic way. Finally in section \ref{conclusions} we give some concluding remarks on the results here presented and we suggest some of their possible applications.

\section{Method} \label{methods}

\subsection{PL structures on a Lie group $G$}

A Poisson-Lie group is a Lie group $G$ together with a Poisson structure $\{,\}$ on $C^\infty(G)$, such that the multiplication 
$\mu:G\otimes G \rightarrow G$ is a Poisson map, namely
\begin{equation}
\{f \circ \mu, g \circ \mu \}_{C^\infty(G \otimes G)} (u,v)= \{f,g\}_{C^\infty(G)}(\mu(u,v)), \qquad u,v \in G, \quad f,g \in C^\infty(G). 
\label{PoissonLiedef}
\end{equation}
In the language of Poisson-Hopf algebras~\cite{CP}, the pull-back of the  multiplication $\mu$ on $G$ defines a coproduct map $\Delta:C^\infty(G) \to C^\infty(G \otimes G)$ through
\begin{equation}
\Delta(f)(u \otimes v)=f(\mu(u,v)) \qquad u,v \in G, \quad f \in C^\infty(G). \label{coproduct}
\end{equation}
In terms of the coproduct map $\Delta$ the homomorphism property (\ref{PoissonLiedef}) can be written in the form
\begin{equation}
\{ \Delta(f), \Delta(g) \}_{C^\infty(G \otimes G)}= \Delta( \{f,g\}_{C^\infty(G)} ).  \label{PoissonLiedef2}
\end{equation}

Our aim is to obtain explicitly and to classify all the (simply connected) real Poisson-Lie groups of dimension three. It is well known that there exist nine non-isomorphic real 3D Lie algebras that cannot be decomposed as a direct sum of lower dimensional real Lie algebras, for which we will follow the structure constants and the basis $\{\We_1,\We_2,\We_3\}$ given in~\cite{pavel}. For any of these Lie algebras $g$, we will select a faithful three dimensional representation $\varrho$ and construct the matrix Lie group element as follows
\begin{displaymath}
M=\exp(z \varrho(\We_1) ) \exp(y \varrho (\We_2) ) \exp(x \varrho(\We_3)). 
\end{displaymath}
Then, we will introduce a set of coordinate functions on $M$ and we use equation (\ref{coproduct}) to define their coproduct. 
For instance, if $X$ is the coordinate function corresponding to the $i,k$ entry of $M$
$$
X(M)=M_{ik},
$$
then from equation (\ref{coproduct}), it follows that the coproduct of $X$ will be given by
\be
\Delta(X)(M \otimes M)=\sum_{j=1}^3 M_{ij} \otimes M_{jk}.
\label{coM}
\ee
Once the coproduct is defined  for the three coordinate functions (denoted by $X,Y,Z$ and expressed in terms of the initial ones $(x,y,z)$),  we will look for the most generic (multiparametric) Poisson bracket for which equation (\ref{PoissonLiedef2}) holds.
Note that if the coordinate functions correspond to linear combinations of the $M_{ij}$ entries, then the natural Ansatz for the Poisson bracket is a quadratic (and obviously antisymmetric) expression in the matrix group entries ($X^1=X,X^2=Y,X^3=Z$) of the form
\begin{equation}
\{ X^\alpha, X^\beta \}(M)= \sum_{i,j,k=1}^3 c^{\alpha \beta}_{ijk} M_{ij} M_{jk}, \qquad \alpha,\beta=1,2,3,  \label{ansatz} 
\end{equation}
where $c^{\alpha \beta}_{ijk}=-c^{\beta \alpha}_{ijk}$ are constant parameters to be determined.
Plugging the coordinate functions into equation (\ref{PoissonLiedef2}) and using the Ansatz (\ref{ansatz}), 
we get a set of linear equations for the coefficients $c^{\alpha \beta}_{ijk}$ that can be easily solved by using a symbolic manipulation program. 
Afterwards we impose the Jacobi identity
\begin{displaymath}
\{ X, \{ Y,Z \} \}+ \{ Z, \{ X, Y \} \}+ \{Y, \{ Z, X \} \} = 0 
\end{displaymath}
and we get a set of quadratic equations for the remaining coefficients $c^{\alpha \beta}_{ijk}$. 
Since the condition (\ref{PoissonLiedef2}) is quite restrictive, it turns out that for the nine groups it is always possible to find the general solution of 
these equations. 

In this way we obtain a multiparametric Poisson bracket  on $C^\infty(G)$ that is both invariant under the coproduct \eqref{coM} and quadratic in the group matrix entries. Now we have to check whether it is the most general one, in the sense that it contains as particular cases all the inequivalent (under group automorphisms) PL structures on $G$, that have to be unambiguously identified. Indeed, since the hypothesys that the Poisson
bracket is quadratic in the group matrix entries is restrictive, this could not be the case. To this aim we use the one-to-one correspondence between PL groups on $G$ and Lie bialgebra structures on $g$ (see \cite{CP}) that we describe in the sequel. 

As we will show, this Ansatz works for seven of the nine three--dimensional real Lie Groups. 
As we will explicitly comment for the remaining two cases, in each case only one non-quadratic term has to be added to the quadratic Ansatz \eqref{ansatz} on the PL bracket in order to get one `lost' PL structure. Apart from that minor variation, the proposed metodology is exactly the same, works for the nine 3D cases and could be used in higher dimensional cases.

\subsection{Classification through the correspondence with Lie bialgebras on $g$}

If $G$ is a Poisson-Lie group with Lie algebra $g$, then it is always possible to define a canonical
Lie algebra structure on $g^\ast$ through
\begin{equation}
[\xi_1,\xi_2]_{g^\ast}=(d\{f_1,f_2\})_e, \label{linearization}
\end{equation}
where $\xi_1,\xi_2 \in g^\ast$, $e$ is the identity element of $G$ and $f_1,f_2 \in C^\infty(G)$ are chosen in such a way that
$(df_i)_e=\xi_i$. This same bracket can be written in the form
\begin{equation}
[\xi_1,\xi_2]_{g^\ast}=\delta^\ast ( \xi_1 \otimes \xi_2), \label{deltastar}
\end{equation}
where $\delta^\ast$ is the dual of a $1-$cocyle of $g$ with values in $g \otimes g$ (the cocommutator)
\begin{equation}
\delta([X,Y])=[\delta(X),Y \otimes 1+ 1 \otimes Y]+[X \otimes 1+ 1 \otimes X, \delta(Y)].  \label{cocycles}
\end{equation}   
The pair $(g,\delta)$ is called the tangent Lie bialgebra of the PL structure on $G$.
We have the following key Theorem, due to Drinfel'd~\cite{Drinfeld}.
\begin{theorem} \label{key}
Let $G$ be a Lie group with Lie algebra $g$. If $G$ is a Poisson-Lie group, then $g$ has a natural Lie bialgebra
structure, called the tangent Lie bialgebra of $G$.
Conversely, if $G$ is connected and simply connected, every Lie bialgebra 
structure on $g$ is the tangent Lie algebra of a unique Poisson structure on $G$ which makes $G$ into a
Poisson-Lie group.
\end{theorem} 
So, in order to identify all PL structures on $G$ we have  to classify  all the possible cocommutators $\delta$ on the Lie algebra $g$, {\em i.e.}, all the tangent Lie bialgebras on $g$.

Again this implies the solution of a set of linear equations (the cocycle condition (\ref{cocycles}))
for a coantisymmetric map $\delta$ with arbitrary coefficients, together with a further quadratic equation on them (the Jacobi coidentity that ensures that $g^\ast$ is a Lie algebra). 
Once this problem has been solved we can compare the Lie algebra structure on $g^\ast$ coming from the cocommutator (\ref{deltastar}) with the one  (\ref{linearization})
coming from the linearization of our quadratic PL bracket. By Theorem \ref{key}, if we find that the latter one is a particular case of the former one, then our Poisson-Lie bracket is not the most general one.
This happens only for two solvable Lie algebras, namely, for $A_{3,4}$ and $A_{3,6}$. In both cases, we find that it suffices to add a single non-quadratic term in order to get a family of generic
Poisson--Lie brackets on the group $G$. Finally, the corresponding Casimir functions for these generic brackets are also explicitly found. 

Afterwards, in order to classify all the so obtained Poisson--Lie structures in equivalence classes under group automorphisms we firstly identify all the coboundary cases. This is done by solving the modified Classical Yang-Baxter equation on the Lie algebra $g$ associated with the Lie group $G$ and by using the components $r^{ij}$ of any of these solutions to construct the Sklyanin bracket
$$
\{f,g\}=r^{ij} \left( X^L_i f X^L_j g -  X^R_i f X^R_j g \right) \qquad f,g \in C^\infty(G),
$$
where $X^L_i, \ X^R_i, \ i=1,2,3$ are, respectively, the left and right invariant vector fields under the group action.
By comparing this Sklyanin bracket with the generic one previusly obtained, we can identify the particular values of the free parameters in our generic PL brackets that give rise to coboundary structures. Obviously, all the remaining ones will be non-coboundaries.
  
Finally, the identification of the equivalence classes of PL structures is done by using Gomez's classification~\cite{gomez} of all unequivalent bialgebra structures (under Lie algebra automorphisms) on the nine real three-dimensional algebras. Indeed, by Theorem \ref{key}, all the Lie algebra structures induced on $g^*$ by the unequivalent cocommutators on $g$ (see eq.~(\ref{deltastar})) found in~\cite{gomez} have to appear as particular cases (corresponding to particular values of the parameters
coming from the full PL structures previously computed) of the linearization of our Poisson-Lie structures on $G$ (see eq.~(\ref{linearization})). Due to the one to one correspondence between PL structures on the Lie group $G$ and
the Lie bialgebra structures on its Lie algebra $g$, a representative of each of the equivalence classes of PL structures on $G$ can be explicitly obtained, and the classification problem is solved.
 
\section{Results} \label{results}
In this section we present in a schematic way the full classification of 3D real PL groups.
We present in more detail the computations for  the solvable Lie algebra $A_{3,2}$  and for the rest of the cases we list the relevant results in the same order and with the same notation.

\subsection{PL structures on the solvable group generated by $A_{3,2}$}

\begin{enumerate}
\item {\bf Commutation relations}. We write (following \cite{pavel}) the commutation relations
of the Lie algebra under scrutiny. 
In particular, the solvable Lie algebra $A_{3,2}$ is defined by the commutation relations 
$$
[\We_1,\We_3]=\We_1, \qquad [\We_2,\We_3]=\We_1+\We_2, \qquad [\We_1,\We_2]=0.
$$

\item {\bf Representation}. We give a faithful three dimensional representation $\varrho$, which in the case of $A_{3,2}$ is given by  
$$
\varrho(\We_1)=\left( 
\begin{array}{ccc}
0 & 0 & 1\\
0 & 0 & 0\\
0 & 0 & 0
\end{array}
\right),  \qquad 
\varrho(\We_2)=\left( 
\begin{array}{ccc}
0 & 0 & 1\\
0 & 0 & 1\\
0 & 0 & 0
\end{array}
\right), \qquad
\varrho(\We_3)=\left( 
\begin{array}{ccc}
-1 & -1 & 0\\
0 & -1 & 0\\
0 & 0 & 0
\end{array}
\right).
$$

\item {\bf Matrix group element}. The matrix group element 
$$
M=\exp(z \varrho(\We_1) ) \exp(y \varrho(\We_2)) \exp(x \varrho(\We_3) )
$$
is computed in terms of the local coordinates on the group given by $(z,y,x)$. When useful, we will define a change of variables by rewriting the $M$-entries using capital letters. 
The aim of this change of variables is to simplify the expressions for the coproduct and for the 
Poisson--Lie brackets (and also to simplify the identification of Poisson brackets that are quadratic in the 
matrix group element entries). The expression for the matrix group element is used to compute the group multiplication law and 
the corresponding coproduct. 

Namely, the matrix group element for $A_{3,2}$ is given by 
$$
M=
\left( 
\begin{array}{ccc}
\exp(-x) & -x \exp(-x) &  y+z\\
0 & \exp(-x) & y\\
0 & 0 & 1
\end{array}
\right)=\left(
\begin{array}{ccc}
 X & \ln(X) X & Z \\
 0 & X & Y\\
 0 & 0 & 1
\end{array} 
\right).
$$ 
Hereafter we will use $X,Y,Z$ as coordinates on the group; since $X=\exp(-x)$, the coordinate $X$ has the constraint $X>0$.  

\item {\bf Coproduct}. It comes from the group multiplication law, and in terms of the coordinate functions is obtained by solving the set of equations ($r={\rm dim}(M)$): 
$$
\Delta(M_{ij})=\sum_{k=1}^r M_{ik} \otimes M_{kj}, \qquad i,j=1,\dots,r.
$$      
for $X,Y,Z$ (here $M_{ij}$ denotes the function that maps the group element $M$ into its $i,j$ entry).

In the case of $A_{3,2}$ we realize that the $X$ coordinate
can be obtained by taking the $M_{11}$ entry, that the $Y$ coordinate is just $M_{23}$ and $Z$ is $M_{13}$. According to the definition (\ref{coproduct}), the coproduct of $X,Y,Z$, evaluated on the two group elements $M_1$ and $M_2$ will be given, respectively, by the $(11)$, $(23)$ and $(13)$ entries of the product $M_1 \cdot M_2$.
Namely:
\begin{eqnarray*}
M_1 \cdot M_2&=& 
\left(
\begin{array}{ccc}
 X_1 & X_1 \ln(X_1) & Z_1 \\
 0 & X_1 & Y_1\\
 0 & 0 & 1
\end{array} 
\right)
\cdot 
\left(
\begin{array}{ccc}
 X_2 & X_2 \ln(X_2) & Z_2 \\
 0 & X_2 & Y_2\\
 0 & 0 & 1
\end{array} 
\right)= \\
&=& \left(
\begin{array}{ccc}
 X_1 X_2 & X_1 X_2 \ln(X_1 X_2) & X_1 Z_2+X_1 \ln(X_1) Y_2 +Z_1 \\
 0 & X_1 X_2 & X_1 Y_2 + Y_1\\
 0 & 0 & 1
\end{array} 
\right)
\end{eqnarray*}
Now by the identification
$$
X_1=X \otimes 1, \quad X_2=1 \otimes X, \quad Y_1=Y \otimes 1, \quad Y_2=1 \otimes Y, 
\quad Z_1=Z \otimes 1, \quad Z_2=1 \otimes Z,
$$
we obtain the coproducts for the coordinate functions
\begin{eqnarray}
\Delta(X)&=&X \otimes X,\nonumber\\
\Delta(Y)&=&X \otimes Y + Y \otimes 1, \label{coa32} \\
\Delta(Z)&=&X \otimes Z+ X \ln(X) \otimes Y+ Z \otimes 1.
\nonumber
\end{eqnarray}

\item {\bf PL brackets}. The most general PL bracket on the group is obtained through the Ansatz (\ref{ansatz}) by solving the equations (\ref{PoissonLiedef2}) for the coproducts of the coordinate functions.  In the $A_{3,2}$ case, the most generic quadratic Poisson bracket for which \eqref{coa32} is a Poisson map is given by the three-parametric structure
 \begin{eqnarray}
\{ X,Y \}_1&=&0, \nonumber \\
\{ X,Z \}_1&=& -a_1 X^2+ b_1X Y +a_1 X, \label{PLA32}\\
\{ Y,Z \}_1&=& c_1(1- X^2)+ \frac{b_1}{2} Y^2 +a_1 Y \nonumber.
\end{eqnarray} 
Note that since one of the group entries is the unity, the quadratic bracket contains linear and constant terms. The Jacobi identity for this bracket is automatically satisfied, and all the possible Lie bialgebra structures on $A_{3,2}$ can be obtained 
as the dual of the linearization of (\ref{PLA32}) (see eq.~(\ref{linearization})). The subindex in the Poisson bracket labels the different possible families of solutions for the PL bracket when the compatibility \eqref{PoissonLiedef2} is imposed. In this case there is only one of such families.

\item {\bf Casimir function}. The Casimir function for (\ref{PLA32}) is found to be
$$
\mathcal{C}= \dfrac{2 c_1 \left(1+X^{2}\right)+Y\left(-2 a_1 (-1+X)+b_1 Y\right)}{ X},
$$
which is real and well defined for all values of the parameters $(a_1,b_1,c_1)$. As we shall see in other groups, the Casimir function could be different depending on the values of the parameters appearing in the generic PL structure.

\item {\bf Coboundary cases}. The most general skewsymmetric candidate for constant classical $r$-matrix on the Lie  algebra $A_{3,2}$ is
\begin{displaymath}
r=r^{12} \, e_1 \wedge e_2 + r^{13} \, e_1 \wedge e_3 +r^{23} \, e_2 \wedge e_3
\end{displaymath}
where $(r^{12},r^{13},r^{23})$ are free real parameters. Now we have to impose that $r$ is a solution of the mCYBE
$$
[ \zeta \otimes 1\otimes 1 + 1\otimes  \zeta \otimes 1 +
1\otimes 1\otimes \zeta,[[r,r]]\, ]=0 \qquad \zeta \in {A_{3,2}}  
$$
and this condition leads to $r^{23}=0$.
Therefore, the coboundary PL structures for $A_{3,2}$ will be generated by
$$
r=r^{12} \, \We_1 \wedge \We_2 +r^{13}  \, \We_1 \wedge \We_3. 
$$ 

In order to identify these structures within \eqref{PLA32} we compute the left and right invariant vector fields for the $A_{3,2}$ group, which read
\begin{eqnarray*}
&\!\!\!\!\!\!\!\! \!\!\!\!\!\!\!\! \quad L_1=L_z=e^{-x} \dfrac{\partial}{\partial x}\qquad\qquad\qquad\qquad
&R_1=R_z=\dfrac{\partial}{\partial x}\\
&\!\!\!\!\!\!\!\! \!\!\!\!\!\!\!\! \qquad L_2=L_y=-x e^{-x} \dfrac{\partial}{\partial x}+e^{-x} \dfrac{\partial}{\partial y}\qquad\qquad
&R_2=R_y=\dfrac{\partial}{\partial y} \\
&\!\!\!\!\!\!\!\! \!\!\!\!\!\!\!\! \!\!\!\!\!\!\!\! L_3=L_x=\dfrac{\partial}{\partial z}\qquad\qquad\qquad\qquad 
&R_3=R_x=-(y+z) \dfrac{\partial}{\partial x} -y  \dfrac{\partial}{\partial y} + \dfrac{\partial}{\partial z}.
\end{eqnarray*}
Using them we can compute the Sklyanin bracket
$$
\{ f, g \}= r^{ij} \left( L_i f L_j g - R_i f R_j g \right)
$$
that for the coordinate functions reads
$$
\{x,y \}=0, \qquad \{x,z \}=r^{13} \left( 1 - \exp(-x) \right), \qquad \{y,z \}=r^{12} \left( 1- \exp(-2x) \right) -r^{13} y.
$$
Passing to the $X,Y,Z$ variables, we obtain:
$$
\{X,Y\}=0, \qquad \{X,Z\}= r^{13} X (X-1), \qquad \{Y,Z\}=  - r^{12} (X^2-1) -r^{13} Y.
$$
Therefore we conclude that (\ref{PLA32}) is a coboundary structure if $b_1=0$, with $a_1=-r^{13}$ and $c_1=r^{12}$. Non-coboundary cases will appear whenever $b_1\neq 0$.

\item {\bf Isomorphism with Lie algebras in~\cite{gomez}}. First of all we have to identify the change of bases in the Lie algebra that  transforms our algebra into the generators  $\Ge_0,\Ge_1,\Ge_2$ used in~\cite{gomez}. In particular, the algebra $A_{3,2}$ is isomorphic to the algebra  $\tau_3'(1)$ in Gomez classification through 
$$
\We_1=\Ge_1 \qquad \We_2=\Ge_2 \qquad \We_3=-\Ge_0.
$$ 

\item {\bf Correspondence with the classification of Lie bialgebras~\cite{gomez}}. 
In the case of $A_{3,2}$ the linearization of the Poisson-Lie bracket (\ref{PLA32}) gives
$$
\{x,y\}=0, \qquad \{x,z\}=-a_1x-b_1y, \qquad \{y,z\}=2c_1x +a_1 y,
$$
where $x,y,z$ are the duals to $\We_3,\We_2,\We_1$ respectively.
The corresponding cocommutator (in the basis~\cite{gomez}) is given by
\begin{eqnarray}
\delta(\Ge_0)&=& - a_1 \, \Ge_0 \wedge \Ge_1 + 2 c_1 \, \Ge_1 \wedge \Ge_2,  \nonumber \\
\delta(\Ge_1)&=& 0 \label{deltaA32}\\
\delta(\Ge_2)&=& - a_1 \, \Ge_1 \wedge \Ge_2 + b_1 \, \Ge_0 \wedge \Ge_1. \nonumber
\end{eqnarray}
By comparing the cocommutator (\ref{deltaA32}) with the four unequivalent classes of Lie Lialgebra structures for $\tau_3'(1)$ found by Gomez, we find
which particular values of the parameters $(a_1,b_1,c_1)$ in (\ref{PLA32}) correspond with each of these classes. Each of these four sets will provide the four inequivalent classes of PL structures on the group generated by $A_{3,2}$.

In general, this correspondence will be summarized in the form of a table. In its first column we write the number that identifies the type of Lie bialgebra (last column of table III in~\cite{gomez}), and the symbol $(\ast)$ is used to distinguish the coboundary Lie bialgebras.
In the second one we give the Poisson bracket $\{,\}_i$ we are considering (in the case of  $A_{3,2}$ we have only one) and in the following columns the particular values of the $(a_i,b_i,c_i)$ parameters for which the linearization of the PL bracket coincides with the Lie bialgebra parameters from~\cite{gomez}. We recall here the assumptions of~\cite{gomez} on the parameters $\alpha,\beta,\lambda,\omega$. All are assumed to be nonzero real numbers, moreover $\alpha$ and $\beta$ can be rescaled (by an appropriate
automorphism of the Lie algebra $g$) to arbitrary nonzero values, $\omega$ can be rescaled to any nonzero value of the same sign, 
and $\lambda$ is an essential parameter. Also the parameters $\rho$ and $\mu$ are subject to the constraints
$-1\leq \rho \leq 1$ and $\mu \geq 0$.
 
In this way, each row in the table corresponds to one of the inequivalent PL structures on the group under consideration, that can be explicitly obtained by substituting the values of the $(a_i,b_i,c_i)$ parameters given in the Table into the generic expression \eqref{PLA32}. In this way, for the $A_{3,2}$ the following inequivalent PL structures are obtained:

\begin{center}
    \begin{tabular}{ c | c | c | c | c |}
   Lie bialg.~in~\cite{gomez} & $\{,\}_i$  & $a_i$ & $b_i$ & $c_i$ \\ 
     \hline
    12 $(\ast)$ & $\{,\}_1$ & $0$ & $0$ & $-\omega$ \\   
    \hline
    (8)\! $(\ast)$ & $\{,\}_1$ & $1$ & $0$ & $0$  \\
    \hline
     13 & $\{,\}_1$ & $0$ & $\lambda$ & $0$  \\
    \hline
     14 & $\{,\}_1$ & $0$ & $\lambda$ & $-\omega$   \\
    \hline
    \end{tabular}
    
    {\bf Table 1}. Classification of PL structures on $A_{3,2}$.
    
\end{center}

\item {\bf Remarks}. Note that  there is {\em not} a one-to-one correspondence between the number  of parameters in the generic PL bracket \eqref{PLA32} and the number of inequivalent classes of PL structures. Also, we mention that there seems to be a misprint in \cite{gomez}, since  in order to get correct expressions for the cocommutators of these Lie bialgebras, the generators $\Ge_1$ and $\Ge_2$ have to be interchanged.

\end{enumerate}

\subsection{PL structures on the Heisenberg group generated by $A_{3,1}$}
\begin{enumerate}
\item Commutation relations
$$
[\We_2,\We_3]=\We_1, \qquad [\We_1,\We_2]=0, \qquad [\We_1,\We_3]=0.
$$
Note that these are also the commutation rules of the massless (1+1) Galilei Lie algebra.

\item Representation
$$
\varrho(\We_1)=\left( 
\begin{array}{ccc}
0 & 0 & 1\\
0 & 0 & 0\\
0 & 0 & 0
\end{array}
\right),  \qquad 
\varrho(\We_2)=\left( 
\begin{array}{ccc}
0 & 1 & 0\\
0 & 0 & 0\\
0 & 0 & 0
\end{array}
\right), \qquad
\varrho(\We_3)=\left( 
\begin{array}{ccc}
0 & 0 & 0\\
0 & 0 & 1\\
0 & 0 & 0
\end{array}
\right).
$$
\item Matrix group element
$$
M=\left(
\begin{array}{ccc}
1 & y & x y+z\\
0 & 1 & x\\
0 & 0 & 1
\end{array}
\right)=\left(
\begin{array}{ccc}
1 & Y & Z\\
0 & 1 & X\\
0 & 0 & 1
\end{array} 
\right).
$$
\item Coproduct
\begin{eqnarray*}
\Delta(X)&=&X \otimes 1 + 1 \otimes X,\\
\Delta(Y)&=&Y \otimes 1 + 1 \otimes Y,\\
\Delta(Z)&=&Z \otimes 1 + 1 \otimes Z+ Y \otimes X.
\end{eqnarray*}
\item Poisson--Lie brackets. In this case we obtain three multiparametric families of Poisson structures that are compatible with the previous coproduct:
\begin{eqnarray*}
\{ X,Y \}_1&=&a_1 X+ b_1 Y,\\
\{ X,Z \}_1&=& \frac{a_1}{2} X^2+ c_1 X+ d_1 Y +b_1 Z,\qquad\qquad\qquad (b_1\neq 0)\\
\{ Y,Z \}_1&=&-\frac{a_1^2 d_1}{b_1^2} X- \frac{b_1}{2} Y^2- \frac{2 a_1 d_1- b_1 c_1}{b_1}Y -a_1 Z.
\end{eqnarray*}
\begin{eqnarray*}
\{ X,Y \}_2&=&0,\\
\{ X,Z \}_2&=& a_2 X+b_2 Y,\\
\{ Y,Z \}_2&=& c_2 X + d_2 Y.
\end{eqnarray*}
\begin{eqnarray*}
\{ X,Y \}_3&=&a_3 X,\\
\{ X,Z \}_3&=& \frac{a_3}{2} X^2+ b_3 X, \qquad\qquad\qquad (a_3\neq 0) \\
\{ Y,Z \}_3&=&-c_3 X+ b_3 Y -a_3 Z.
\end{eqnarray*}

\item Casimir function. Note that for the structures 2 and 3 it depends on the values of the corresponding parameters, that are indicated between parenthesis.
$$
\mathcal{C}_{1}=\dfrac{2( b_1 c_1 - a_1 d_1) X+b_1 ^{2}(2 Z-XY)-2 d_1 (a_1 X+b_1 Y)\log{(a_1 X+b_1 Y)}}
{(a_1 X+b_1 Y)}
$$
$$
\mathcal{C}_{2}= \left\{ 
\begin{array}{l} 
\displaystyle \left[\frac{(\alpha+a_2-d_2) Y-2 c_2 X}{(\alpha-a_2+d_2) Y+2 c_2 X}\right]^{a_2+d_2} \left[
\left( \frac12 (a_2-d_2) Y -c_2 X \right)^2
-\frac{\alpha^2}{4} Y^2\right]^\alpha \\
(\alpha=\sqrt{(a_2-d_2)^2+4 b_2 c_2}, \quad (a_2-d_2)^2+4 b_2 c_2>0) \\ 
\\ 
\displaystyle (a_2+d_2) \arctan\left(\frac{2 c_2 X-(a_2-d_2) Y}{\alpha Y} \right) +\frac12 \alpha \ln\left[\frac{\alpha^2}{4}Y^2+
\left( \frac{a_2-d_2}{2} Y -c_2 X \right)^2 \right] \\
(\alpha=\sqrt{-(a_2-d_2)^2-4 b_2 c_2}, \quad (a_2-d_2)^2+4 b_2 c_2<0) \\ \\
\displaystyle \frac{\exp\left(  \displaystyle \frac{2 (a_2+d_2) c_2 X}{(a_2-d_2)((d_2-a_2) Y+2c_2 X)}\right)}{2 c_2 X+(d_2-a_2) Y} \quad
(c_2 \neq 0, \quad b_2=-(a_2-d_2)^2/(4c_2),  \quad a_2 \neq d_2 ) \\  \\
\displaystyle X \exp\left(- \frac{a_2 Y}{c_2 X}  \right) \quad (d_2=a_2,  \quad b_2 =0, \quad c_2 \neq 0)\\ \\
\displaystyle Y \exp\left(- \frac{a_2 X}{b_2 Y}  \right) \quad (a_2 =d_2,  \quad c_2 =0, \quad b_2 \neq 0) \\ \\
\displaystyle \frac{X}{Y}  \quad (a_2=d_2,  \quad b_2=c_2 =0) 
\end{array} 
\right. 
$$
$$
\mathcal{C}_{3} = \left\{
\begin{array}{l}
X \exp \left[  \displaystyle \dfrac{ 2 b_3 Y+a_3 (XY-2  Z)}{2 c_3 X}  \right] \qquad (c_3 \neq 0) \\
\displaystyle \dfrac{2 b_3 Y+a_3 (XY-2  Z)}{X}  \qquad (c_3=0) 
\end{array}
\right.
$$
\item Coboundary cases. The most generic classical $r$-matrix on this algebra is
$$
 r=r^{12} \, e_1 \wedge e_2 + r^{13} \, e_1 \wedge e_3 + r^{23} \, e_2 \wedge e_3.
$$
The Sklyanin bracket associated to this $r$ is just the PL bracket $\{ \,, \,\}_2$ provided that 
\begin{displaymath}
a_2=d_2=-r^{23}, \qquad b_2=c_2=0.  
\end{displaymath}
(coefficients $r^{12}$ and $r^{13}$ do not play any role in the above Poisson bracket) .
Therefore, the PL brackets $\{ \,, \,\}_1$ and $\{ \,, \,\}_3$ are always non-coboundary ones. 

\item Isomorphism with the  $\mathfrak{n}_3$ Lie algebra in~\cite{gomez}:
$$
\We_1=\Ge_0 \qquad \We_2=\Ge_1 \qquad \We_3=\Ge_2.
$$ 
\item Correspondence with the classification of Lie bialgebras:

\begin{center}
    \begin{tabular}{ c | c | c | c | c | c |}
    Lie bialg.~in~\cite{gomez} & $\{,\}_i$  & $a_i$ & $b_i$ & $c_i$ & $d_i$ \\ 
     \hline
    (5-5')& $\{,\}_2$ & $-\rho$ & $0$ & $0$ & $-1$ \\   
    \hline
    (12) & $\{,\}_2$ & $-1$ & $0$ & $1$ &  $-1$  \\
    \hline
     (15) & $\{,\}_2$ & $-\mu$ & $1$ & $1$ &  $-\mu$  \\
    \hline
    \multirow{1}{*}{17} & $\{,\}_2$ & $0$ & $0$ & $1$ & $0$ \\
                           \hline
     (13) & $\{,\}_3$ & $-1$ & $0$ & $\lambda$ &  $\cdot$  \\
    \hline
       (10) & $\{,\}_3$ & $-1$ & $0$ & $0$ &  $\cdot$  \\
    \hline
    \end{tabular}
        
    {\bf Table 2}. Classification of PL structures on $A_{3,1}$.

\end{center}
\item Remarks. A first study of the PL structures on the Heisenberg group and their quantizations was performed in~\cite{Kuper} and Heisenberg Lie bialgebras together with their corresponding quantum algebras were presented in~\cite{BHPheis}. The classification and construction of Heisenberg PL groups given in~\cite{galileo} is included in Table 2. The only coboundary structure corresponds to the `isolated point' (5-5') with $\rho=1$. It can be also cheked that the PL brackets $\{,\}_1$ and $\{,\}_3$ are isomorphic for generic values of the parameters, which explains the absence of  $\{,\}_1$ in Table 2.

\end{enumerate}

\subsection{PL structures on the `book group' generated by $A_{3,3}$} \label{qbook}
\begin{enumerate}
\item Commutation relations
$$
[\We_1,\We_3]=\We_1, \qquad [\We_2,\We_3]=\We_2, \qquad [\We_1,\We_2]=\We_0.
$$
\item Representation
$$
\varrho(\We_1)=\left( 
\begin{array}{ccc}
0 & 0 & 1\\
0 & 0 & 0\\
0 & 0 & 0
\end{array}
\right),  \qquad 
\varrho(\We_2)=\left( 
\begin{array}{ccc}
0 & 0 & 0\\
0 & 0 & 1\\
0 & 0 & 0
\end{array}
\right), \qquad
\varrho(\We_3)=\left( 
\begin{array}{ccc}
-1 & 0 & 0\\
0 & -1 & 0\\
0 & 0 & 0
\end{array}
\right).
$$
\item Matrix group element
$$
M=
\left( 
\begin{array}{ccc}
\exp(-x) & 0 & z\\
0 & \exp(-x) & y\\
0 & 0 & 1
\end{array}
\right)=
\left(
\begin{array}{ccc}
 X & 0 & Z \\
 0 & X & Y\\
 0 & 0 & 1
\end{array} 
\right), \qquad X>0.
$$ 
\item Coproduct
\begin{eqnarray*}
\Delta(X)&=&X \otimes X,\\
\Delta(Y)&=&X \otimes Y + Y \otimes 1,\\
\Delta(Z)&=&X \otimes Z + Z \otimes 1.
\end{eqnarray*}
\item Poisson--Lie brackets
\begin{eqnarray*}
\{ X,Y \}_1&=& a_1 (X^2-X)-b_1 X Y-2 c_1 X Z,\\
\{ X,Z \}_1&=& d_1 (X^2-X)+2 e_1 X Y+b_1 X Z,\\
\{ Y,Z \}_1&=& f_1 (1-X^2)+e_1 Y^2+b_1 Y Z-d_1 Y+c_1 Z^2+a_1 Z.
\end{eqnarray*}
\item Casimir
$$
\mathcal{C}=\dfrac{1}{X}\left(f_1(1+X^{2})+d_1(-1+X)Y+e_1 Y^{2}+a_1 Z(1-X)+Z(b_1 Y+c_1 Z)\right).
$$
\item Coboundary cases. The most generic classical $r$-matrix  is now
$$
r=r^{12} \, \We_1 \wedge \We_2+ r^{13} \, \We_1 \wedge \We_3 +r^{23} \, \We_2 \wedge \We_3,
$$
which corresponds to the PL brackets given by
$$
a_1=r^{23}, \qquad d_1=r^{13}, \qquad f_1=r^{12}, \qquad b_1=c_1=e_1=0.
$$

\item Isomorphic to the algebra  $\tau_3(1)$ in~\cite{gomez} through the change of variables
$$
\We_1=\Ge_1 \qquad \We_2=\Ge_2 \qquad \We_3=-\Ge_0.
$$ 
\item Correspondence with the classification of Lie bialgebras:
\begin{center}
    \begin{tabular}{ c | c | c | c | c | c | c | c |}
    Lie bialg.~in~\cite{gomez} & $\{,\}_i$ & $a_i$ & $b_i$ & $c_i$ & $d_i$ & $e_i$ & $f_i$ \\ 
     \hline
    $5$ ($\rho=1$) $(\ast)$ & $\{,\}_1$ & $0$ & $0$ & $0$ & $0$ & $0$ & $-1$ \\   
    \hline
    $6$ ($\rho=1$, $\chi=\Ge_0 \wedge \Ge_1$) $(\ast)$ & $\{,\}_1$ & $0$ & $0$ & $0$ & $-1$ & $0$ & $0$ \\   
    \hline
    $7$  ($\rho=1$) & $\{,\}_1$ & $0$ & $\lambda$ & $0$ & $0$ & $0$ & $0$ \\   
    \hline
    $(1)$ & $\{,\}_1$ & $0$ & $\lambda$ & $0$ & $0$ & $0$ & $-\alpha$ \\   
    \hline
    $(2)$ & $\{,\}_1$ & $0$ & $0$ & $\lambda/2$ & $0$ & $\lambda/2$ & $-\omega$ \\
     \hline
    $9$ & $\{,\}_1$ & $0$ & $0$ & $\lambda/2$ & $0$ & $\lambda/2$ & $0$ \\ 
    \hline 
    $10$ & $\{,\}_1$ & $0$ & $0$ & $-1/2$ & $0$ & $0$ & $0$ \\ 
    \hline 
    $11$ & $\{,\}_1$ & $0$ & $0$ & $-1/2$ & $0$ & $0$ & $-\omega$ \\ 
    \hline 
    $(3)$ & $\{,\}_1$ & $0$ & $0$ & $-1/2$ & $-\alpha$ & $0$ & $0$ \\ 
    \hline 
    \end{tabular}

    {\bf Table 3}. Classification of PL structures on $A_{3,3}$.

\end{center}
\item Remarks. Note that in Table 3 the parameter $a$ is always $0$. This is due to the fact that $a$ is equivalent to $d$ under the Lie group automorphism  $Y \leftrightarrow Z$. This classification of PL structures on the book group has been recently presented in~\cite{BBMbook}, where it has been explicitly shown that many of these structures correspond -under suitable changes of local coordinates- to Poisson versions of 3D quantum algebras (see also~\cite{Marmo}, in which the PL group corresponding to (1)  with $b_1=1/f_1$ was constructed). Also, the PL structure given by the case 7 has been recently shown to underlie the integrability of a class of 3D Lotka-Volterra systems (see~\cite{BBMbook} and references therein).

\end{enumerate}

\subsection{PL structures on the (1+1) Poincar\'e group generated by  $A_{3,4}$}
\begin{enumerate}
\item Commutation relations
$$
[\We_1,\We_3]=\We_1, \qquad [\We_2,\We_3]=-\We_2, \qquad [\We_1,\We_2]=0.
$$
\item Representation
$$
\varrho(\We_1)=\left( 
\begin{array}{ccc}
0 & 0 & 1\\
0 & 0 & 0\\
0 & 0 & 0
\end{array}
\right),  \qquad 
\varrho(\We_2)=\left( 
\begin{array}{ccc}
0 & 0 & 0\\
0 & 0 & -1\\
0 & 0 & 0
\end{array}
\right), \qquad
\varrho(\We_3)=\left( 
\begin{array}{ccc}
-1 & 0 & 0\\
0 & 1 & 0\\
0 & 0 & 0
\end{array}
\right).
$$
\item Matrix group element
$$
M=
\left( 
\begin{array}{ccc}
\exp(-x) & 0 & z\\
0 & \exp(x) & -y\\
0 & 0 & 1
\end{array}
\right)=\left(
\begin{array}{ccc}
 X & 0 & Z \\
 0 & X^{-1} & Y\\
 0 & 0 & 1
\end{array} 
\right), \qquad X>0.
$$
\item Coproduct
\begin{eqnarray*}
\Delta(X)&=&X \otimes X,\\
\Delta(Y)&=&X^{-1} \otimes Y + Y \otimes 1,\\
\Delta(Z)&=&X \otimes Z + Z \otimes 1.
\end{eqnarray*}
\item Poisson--Lie brackets. We have two different families
\begin{eqnarray*}
\{ X,Y \}_1&=&-a_1 X Y+b_1 (X-1),\\
\{ X,Z \}_1&=&c_1 (X-X^2)-a_1 X Z,\\
\{ Y,Z \}_1&=&a_1 Y Z-c_1 Y-b_1 Z.
\end{eqnarray*}
\begin{eqnarray*}
\{ X,Y \}_2&=&a_2 (1- X),\\
\{ X,Z \}_2&=& b_2 (X^2- X), \qquad\qquad\qquad (c_2\neq 0)\\
\{ Y,Z \}_2&=& b_2 Y+a_2 Z+c_2 \ln(X).
\end{eqnarray*}
\item Casimirs
$$
\mathcal{C}_{1}=\left\{
\begin{array}{l} 
\displaystyle \dfrac{c_1(X-1)+a_1 Z}{b_1(1- X)+a_1 XY} \qquad (a_1 \neq 0)\\ \\
\displaystyle \dfrac{b_1 Z+c_1 XY}{X-1} \qquad (a_1=0)
\end{array}
\right.
$$
$$
\mathcal{C}_{2}=
\displaystyle \exp\left[ \displaystyle \dfrac{b_2 X Y+a_2 Z}{c_2 (X-1)}\right]\left( \displaystyle \dfrac{X^{\frac{X}{X-1}}}{X-1}\right) 
$$
\item Coboundary cases. The generic classical $r$-matrix is
$$
 r=r^{12} \, \We_1 \wedge \We_2 + r^{13} \, \We_1 \wedge \We_3 + r^{23} \, \We_2 \wedge \We_3 ,
$$ 
which means that the bracket $\{,\}_1$ is coboundary for the following values of the parameters:
$$
a_1=0, \qquad b_1=r^{23}, \qquad c_1=-r^{13}.
$$ 
(the coefficient $r^{12}$ do not play any role in the above Poisson bracket) .
The bracket $\{,\}_2$ is always a non-coboundary one.

\item Isomorphic to the algebra  $\tau_3(-1)$ in~\cite{gomez} through the change of variables
$$
\We_1=\Ge_1 \qquad \We_2=\Ge_2 \qquad \We_3=-\Ge_0.
$$ 
\item Correspondence with the classification of Lie bialgebras:
\begin{center}
    \begin{tabular}{ c | c | c | c | c |}
    Lie bialg.~in~\cite{gomez} & $\{,\}_i$ & $a_i$ & $b_i$ & $c_i$ \\ 
     \hline
    6 ($\rho=-1$, $\chi=\Ge_0 \wedge \Ge_1$) $(\ast)$ & $\{,\}_1$ & $0$ & $0$ & $1$ \\   
    \hline
    7 ($\rho=-1$) & $\{,\}_1$ & $-\lambda$ & $0$ & $0$ \\   
      \hline
    \multirow{1}{*}{(11) $(\ast)$} & $\{,\}_1$ & $0$ & $-\alpha \beta$ & $\alpha$ \\
 \hline
    5' & $\{,\}_2$ & $0$ & $0$ & $1$ \\   
    \hline
    8 & $\{,\}_2$ & $-\alpha$ & $0$ & $1$ \\   
    \hline
    (14) & $\{,\}_2$ & $\alpha \lambda$ & $-\alpha$ & $1$ \\   
     \hline
    
    \end{tabular}
      
    {\bf Table 4}. Classification of PL structures on $A_{3,4}$.
    
\end{center}
\item Remarks. The $A_{3,4}$ algebra is isomorphic to the (1+1) Poincar\'e algebra written in a `null-plane' basis. Note that the Poisson bracket $\{,\}_2$ is not quadratic in the group matrix entries since it contains a term of the type $c \ln(X)$, that we have been forced to include if we want to recover the non-coboundary PL structures $5'$, $8$ and $(14)$. If we write both families of Poisson brackets in the local coordinates $(x,y,z)$ we obtain
\begin{eqnarray*}
\left\{x,y\right\}_1&=& a_1 y+b_1 (1-\exp(x)) \\
\left\{x,z\right\}_1&=& c_1 (\exp(-x)-1)+a_1 z \\
\left\{y,z\right\}_1&=& y(a_1 z-c_1) +b_1 z
\end{eqnarray*}
\begin{eqnarray*}
\left\{x,y\right\}_2&=& a_2 (\exp(x)-1) \\
\left\{x,z\right\}_2&=& b_2 (1-\exp(-x)) \\
\left\{y,z\right\}_2&=& c_2 x+ b_2 y-a_2 z
\end{eqnarray*}
where the $\left\{y,z\right\}_i$ bracket would be the Poisson version for the
non-commutative (null-plane) Minkowski spacetime associated to the corresponding quantum Poincar\'e group (see~\cite{BHPOS} and references therein).

\end{enumerate}

\subsection{PL structures on the solvable group generated by $A_{3,5}$}

\begin{enumerate}
\item Commutation relations
$$
[\We_1,\We_3]=\We_1, \qquad [\We_2, \We_3]= \rho \, \We_2, \qquad [\We_1, \We_2]=0, \qquad  0< |\rho| <1.
$$
\item Representation
$$
\varrho(\We_1)=\left( 
\begin{array}{ccc}
0 & 0 & 1\\
0 & 0 & 0\\
0 & 0 & 0
\end{array}
\right),  \qquad 
\varrho(\We_2)=\left( 
\begin{array}{ccc}
0 & 0 & 0\\
0 & 0 & \rho\\
0 & 0 & 0
\end{array}
\right), \qquad
\varrho(\We_3)=\left( 
\begin{array}{ccc}
-1 & 0 & 0\\
0 & -\rho & 0\\
0 & 0 & 0
\end{array}
\right).
$$
\item Matrix group element
$$
M=\left( 
\begin{array}{ccc}
\exp(-x) & 0 & z\\
0 & \exp(- \rho x) & \rho y\\
0 & 0 & 1
\end{array}
\right)=\left(
\begin{array}{ccc}
 X & 0 & Z \\
 0 & X^\rho & Y\\
 0 & 0 & 1
\end{array} 
\right), \qquad X>0.
$$
\item Coproduct
\begin{eqnarray*}
\Delta(X)&=&X \otimes X,\\
\Delta(Y)&=&X^\rho \otimes Y+Y \otimes 1,\\
\Delta(Z)&=&X \otimes Z + Z \otimes 1.
\end{eqnarray*}
\item Poisson--Lie brackets. We have three different families:

\begin{eqnarray*}
\{ X,Y \}_1&=& -a_1 X Y+b_1 X (X^{\rho}-1), \\
\{ X,Z \}_1&=& c_1 (X- X^2)+\frac{a_1}{\rho} X Z \qquad \qquad (a_1 \neq 0),\\
\{ Y,Z \}_1&=&\frac{\rho \, b_1 c_1}{a_1} (1-X^{1+\rho})+a_1 Y Z+\rho \, c_1 Y+b_1 Z.
\end{eqnarray*}
\begin{eqnarray*}
\{ X,Y \}_2&=&0,\\
\{ X,Z \}_2&=&  a_2 (X-X^2),\\
\{ Y,Z \}_2&=& b_2(1- X^{1+\rho})+\rho \, a_2 Y.
\end{eqnarray*}
\begin{eqnarray*}
\{ X,Y \}_3&=& a_3 X (X^{\rho}-1),\\
\{ X,Z \}_3&=& 0 \qquad \qquad \qquad  \qquad \qquad \qquad (a_3 \neq 0),\\
\{ Y,Z \}_3&=& b_3(1- X^{1+\rho})+a_3 Z.
\end{eqnarray*}
\item Casimirs
$$
\mathcal{C}_{1}= X^{-\rho}\left(b_1(1-X^{\rho})+a_1 Y\right)(\rho \, c_1 (X-1)-a_1 Z)^{\rho}.
$$
$$
\mathcal{C}_{2}= \left(1-\dfrac{1}{X}\right)^{\rho}\left(b_2(1-X^{\rho})+\rho \, a_2 Y\right).
$$
$$
\mathcal{C}_{3}= (X^{-\rho}-1) \left( b_3 (X-1)-a_3 Z \right)^\rho.
$$
\item Coboundary cases. The bracket $\{,\}_2$ is coboundary for the following values of the parameters and $r$-matrix:
$$
a_2=-r^{13}, \qquad b_2=\rho \, r^{12}, \qquad r=r^{12} \, \We_1 \wedge \We_2+ r^{13}\, \We_1 \wedge \We_3.
$$
The bracket $\{,\}_3$ is coboundary  when
$$
a_3= \rho \, r^{23}, \qquad b_3=\rho \, r^{12}, \qquad r=r^{12} \, \We_1 \wedge \We_2+ r^{23}\, \We_2 \wedge \We_3.
$$
The bracket $\{,\}_1$ is always a non-coboundary one. 

\item Isomorphic to the algebra  $\tau_3(\rho)$, $0< |\rho|<1$ in~\cite{gomez} through the change of basis
$$
\We_1=\Ge_1 \qquad \We_2=\Ge_2 \qquad \We_3=-\Ge_0.
$$ 
\item Correspondence with the classification of Lie bialgebras:
\begin{center}
    \begin{tabular}{ c | c | c | c | c |}
    Lie bialg.~in~\cite{gomez} & $\{,\}_i$ & $a_i$ & $b_i$ & $c_i$ \\ 
     \hline
    \multirow{1}{*}{5} $(\ast)$ & $\{,\}_2$ & $0$ & $-\rho$ & $\cdot$ \\   
    \hline
     6 ($\chi=\Ge_0 \wedge \Ge_1$) $(\ast)$ & $\{,\}_2$ & $1$ & $0$ & $\cdot$  \\
    \hline
    7 & $\{,\}_1$ & $\lambda \,  \rho$ & $0$ & $0$  \\
    \hline
    \end{tabular}
      
    {\bf Table 5}. Classification of PL structures on $A_{3,5}$.
    
\end{center}
\item Remarks. To the best of our knowledge, this case has not been considered in the literature so far. Note that the non-coboundary PL group corresponding to the Lie bialgebra 7 is one of the Lotka-Volterra brackets~\cite{LV} that generalize the PL bracket previously obtained on the book group $A_{3,3}$.
\end{enumerate}

\subsection{PL structures on the 2D Euclidean group generated by  $A_{3,6}$}
\begin{enumerate}
\item Commutation relations
$$
[\We_1,\We_3]=-\We_2, \qquad [\We_2, \We_3]= \We_1, \qquad [\We_1, \We_2]= 0.
$$
\item Representation
$$
\varrho(\We_1)=\left( 
\begin{array}{ccc}
0 & 0 & 0\\
0 & 0 & -1\\
0 & 0 & 0
\end{array}
\right),  \qquad 
\varrho(\We_2)=\left( 
\begin{array}{ccc}
0 & 0 & 1\\
0 & 0 & 0\\
0 & 0 & 0
\end{array}
\right), \qquad
\varrho(\We_3)=\left( 
\begin{array}{ccc}
0 & -1 & 0\\
1 & 0 & 0\\
0 & 0 & 0
\end{array}
\right).
$$
\item Matrix group element
$$
M=
\left( 
\begin{array}{ccc}
\cos(x) & -\sin(x) & y\\
\sin(x) & \cos(x) & -z\\
0 & 0 & 1
\end{array}
\right)=\left(
\begin{array}{ccc}
C & -S & Y\\
S & C & Z\\
0 & 0 & 1
\end{array} 
\right), \qquad C^2+S^2=1.
$$
\item Coproduct
\begin{eqnarray*}
\Delta(C)&=&C \otimes C- S \otimes S,\\
\Delta(S)&=& S \otimes C+ C \otimes S,\\
\Delta(Y)&=&C \otimes Y- S \otimes Z +Y \otimes 1,\\
\Delta(Z)&=&S \otimes Y + C \otimes Z + Z \otimes 1.
\end{eqnarray*}
\item Poisson--Lie brackets
\begin{eqnarray*}
\{ C,S \}_1&=& 0,  \\
\{ C, Y \}_1&=&a_1(1-C^2) +b_1 S (1-C)-c_1 S Y,\\ 
\{ C,Z \}_1&=& a_1 S (1-C)+b_1 (C^2-1)-c_1 S Z,\\
\{ S,Y \}_1&=&-a_1 C S+b_1 (C^2- C)+c_1 C Y,\\
\{ S,Z \}_1&=&a_1 (C^2-C)+ b_1 C S+c_1 C Z,\\
\{ Y,Z \}_1&=&a_1 Z+b_1 Y -\frac{c_1}2( Y^2 +Z^2).
\end{eqnarray*}
\begin{eqnarray*}
\{ C,S \}_2&=& 0, \\
\{ C,Y \}_2&=& a_2 (1-C^2) +b_2 S (1-C),\\ 
\{ C,Z \}_2&=& a_2 S (1-C) +b_2(C^2-1)  \qquad \qquad \qquad (c_2 \neq 0),\\
\{ S,Y \}_2&=&-a_2 C S+b_2 (C^2- C),\\
\{ S,Z \}_2&=& a_2 (C^2-C)+ b_2 C S,\\
\{ Y,Z \}_2&=& a_2 Z+b_2 Y+ c_2 \arccos(C).
\end{eqnarray*}
\item Casimirs
$$
\mathcal{C}_1=\left \{ 
\begin{array}{l} 
2 \arctan\left(\dfrac{c_1 Z-a_1 (1-C)+b_1 S}{b_1 (1-C)+a_1 S-c_1 Y}\right)-\arctan\left(\dfrac{C}{S}\right) \qquad (c_1 \neq 0),\\
\\
a_1 Y-b_1 Z +\dfrac{S \left(a_1 Z+b_1 Y \right)}{C-1} \qquad (c_1=0). 
\end{array}
\right.
$$
$$
\mathcal{C}_2=c_2 \ln \left( 1- C \right) +a_2 Y-b_2 Z +\frac{S \left(a_2 Z+b_2 Y+ c_2 \arccos(C)\right)}{C-1}
$$
\item Coboundary cases. The generic classical $r$-matrix is given by
$$
r=r^{12} \, \We_1 \wedge \We_2 + r^{13} \, \We_1 \wedge \We_3 +r^{23} \, \We_2 \wedge \We_3.
$$
The bracket $\{,\}_1$ is coboundary for the following values of the parameters 
$$
a_1=r^{13}, \qquad b_1=-r^{23}, \qquad c_1=0,
$$
(coefficient $r^{12}$  do not play any role in the above Poisson bracket) 
while the bracket $\{,\}_2$ is always a non-coboundary one.

\item Isomorphic to the algebra  $s_3(0)$ in~\cite{gomez} through the change of basis
$$
\We_1=\Ge_1 \qquad \We_2=\Ge_2 \qquad \We_3=-\Ge_0.
$$ 
\item Correspondence with the classification of Lie bialgebras:
\begin{center}
    \begin{tabular}{ c | c | c | c | c |}
    Lie bialg.~in~\cite{gomez} & $\{,\}_i$ & $a_i$ & $b_i$ & $c_i$ \\ 
     \hline
   (9)  & $\{,\}_1$ & $0$ & $0$ & $-\lambda$ \\  
   \hline 
   15'  & $\{,\}_2$ & $0$ & $0$ & $-\omega$ \\
    \hline
   \multirow{1}{*}{(11') $(\ast)$}  & $\{,\}_1$ & $-1$ & $0$ & $0$  \\
              \hline
   (14') & $\{,\}_2$ & $-\alpha$& $0$ & $-\lambda$  \\
    \hline
    \end{tabular}
      
    {\bf Table 6}. Classification of PL structures on $A_{3,6}$.
    
\end{center}
\item Remarks. In this case the Poisson bracket $\{,\}_2$ is not quadratic in the group matrix entries since it contains a term $\arccos(C)$, that has been allowed in order to obtain the PL structure corresponding to the Lie bialgebras $(14')$ and $15'$. PL structures on the euclidean group were firstly studied in~\cite{euclideo}, where the case (14') is lacking. 

\end{enumerate}

\subsection{PL structures on the solvable Lie group generated by  $A_{3,7}$}

\begin{enumerate}
\item Commutation relations
$$
[\We_1,\We_3]=\mu \We_1-\We_2, \qquad [\We_2, \We_3]= \We_1+\mu \We_2, \qquad [\We_1, \We_2]= 0, \qquad \mu>0.
$$
\item Representation
$$
\varrho(\We_1)=\left( 
\begin{array}{ccc}
0 & 0 & \mu\\
0 & 0 & -1\\
0 & 0 & 0
\end{array}
\right),  \qquad 
\varrho(\We_2)=\left( 
\begin{array}{ccc}
0 & 0 & 1\\
0 & 0 & \mu\\
0 & 0 & 0
\end{array}
\right), \qquad
\varrho(\We_3)=\left( 
\begin{array}{ccc}
-\mu & -1 & 0\\
1 & -\mu & 0\\
0 & 0 & 0
\end{array}
\right).
$$
\item Matrix group element
\begin{eqnarray*}
&& M=
\left( 
\begin{array}{ccc}
e^{-\mu x} \cos(x) & -e^{-\mu x} \sin(x) & y+ \mu z\\
e^{-\mu x} \sin(x) & e^{-\mu x} \cos(x) & \mu y -z\\
0 & 0 & 1
\end{array}
\right)
=\left(
\begin{array}{ccc}
C & -S & Y\\
S & C & Z\\
0 & 0 & 1
\end{array} 
\right), \\
&& C^2+S^2-\exp \left( -2 \mu \arctan \left( \frac{S}C \right) \right)=0. 
\end{eqnarray*}
\item Coproduct
\begin{eqnarray*}
\Delta(C)&=&C \otimes C- S \otimes S,\\
\Delta(S)&=& S \otimes C+ C \otimes S,\\
\Delta(Y)&=&C \otimes Y- S \otimes Z +Y \otimes 1,\\
\Delta(Z)&=&S \otimes Y + C \otimes Z + Z \otimes 1.
\end{eqnarray*}
\item Poisson--Lie brackets
\begin{eqnarray*}
\{ C,S \}_1&=& 0 \\
\{ C, Y \}_1&=&\dfrac{1}{2a_1} (\mu C+ S)\left[c_1^{2}(1+\mu^{2})(-Y+\mu Z)+ b_1 (-2 a_1 S+b_1 (-Y+\mu Z))+ \right.\\
            && +\left. 2c_1 [a_1(-1 +C+\mu S)+b_1 \mu (Y-\mu Z)]\right],\\ 
\{ C,Z \}_1&=& -\dfrac{1}{2 a_1}(\mu C+ S) \left[ b_1^{2}(\mu Y+Z)-2 b_1  \left[a_1 (C-1)+c_1 \mu (\mu Y+ Z)\right]+ \right.\\
           &&  + \left. c_1 \left[2 a_1 \mu (C -1)-2 a_1 S+c_1 (1+\mu^{2})(\mu Y+Z)\right]\right],\\
\{ S,Y \}_1&=&-\dfrac{1}{2 a_1}(C\!-\!\mu S)\left[c_1^{2}(1+\mu^{2})(-Y+\mu Z)+b_1 (-2 a_1 S+b_1 (-Y+\mu Z))+ \right. \\
           && + \left. 2 c_1 \left[a_1 (-1+C+\mu S)+b_1 \mu (Y-\mu Z)\right]\right],\\
\{ S,Z \}_1&=&\dfrac{1}{2a_1}(C-\mu S)\left[b_1^{2}(\mu Y+Z)-2b_1 \left[a_1(C-1)+c_1 \mu (\mu Y+Z)\right] + \right. \\
           && + \left. c_1 \left[2 a_1 \mu (C-1)-2 a_1 S +c_1 (1+\mu^{2})(\mu Y+Z)\right]\right],\\
\{ Y,Z \}_1&=&-\dfrac{1}{4a_1}\left[-4a_1^{2}(C^{2}+S^{2}-1)+4 a_1 [(c_1+\mu b_1 -c_1 \mu^{2})Y+ (b_1-2 c_1 \mu )Z]  \right.+ \\
           && + \left. (1+\mu^{2})(c_1^{2}+(b_1-c_1 \mu)^{2})(Y^{2}+Z^{2})\right], \\
 (a_1 &\neq& 0).
\end{eqnarray*}
\begin{eqnarray*}
\{ C,S \}_2&=& 0,  \\
\{ C,Y \}_2&=&  \frac{a_2 (Y-\mu Z) (\mu C+S)}{\mu},\\ 
\{ C,Z \}_2&=&  \frac{a_2 (\mu Y+Z) (\mu C+S)}{\mu},\\
\{ S,Y \}_2&=& -\frac{a_2 (Y-\mu Z) (C-\mu S )}{\mu},\\
\{ S,Z \}_2&=& -\dfrac{a_2(\mu Y+Z)(C-\mu S)}{\mu},\\
\{ Y,Z \}_2&=& -\frac{a_2 (\mu^2+1) (Y^2+Z^2)}{2 \mu}.
\end{eqnarray*}
\item Casimirs
\begin{eqnarray*}
\mathcal{C}_{1}&=&\arctan\left(\dfrac{S}{C}\right)+\dfrac{1}{(i+\mu)}\ln\left[
\dfrac{-2 i a_1 (-1 +C-i S)+(i+\mu)(i b_1 +c_1 -i c_1 \mu)(Y-i Z)}
{(i+\mu)(b_1-c_1 (i+\mu))}
\right]\\	
&& +\dfrac{1}{(-i+\mu)}\ln\left[
\dfrac{2 i a_1 (-1 +C+i S)+(\mu+i)(i b_1 +c_1 +i c_1 \mu)(Y+i Z)}
{(\mu-i)(b_1+c_1 (i-\mu))}
\right].\\
\mathcal{C}_2&=&\arctan{\left(\dfrac{S}{C}\right)}+\dfrac{2}{1+\mu^{2}} 
\left[\arctan\left(\dfrac{Z}{Y}\right)-\dfrac{\mu}{2}\ln (Y^{2}+Z^{2})\right].
\end{eqnarray*}
\item Coboundary cases. For this algebra the most general classical $r$-matrix is
$$
r=r^{12} \, \We_1 \wedge \We_2,
$$
which corresponds to the bracket $\{,\}_1$ for the following values of the parameters
$$
a_1=r^{12}(1+\mu^2), \qquad b_1=c_1=0.
$$
The bracket $\{,\}_2$ is always a  non-coboundary one.

\item Isomorphic to the algebra  $s_3(\mu)$ in~\cite{gomez} through the change of basis
$$
\We_1=\Ge_1 \qquad \We_2=\Ge_2 \qquad \We_3=-\Ge_0.
$$ 
\item Correspondence with the classification of Lie bialgebras:
\begin{center}
    \begin{tabular}{ c | c | c | c | c |}
    Lie bialg.~in~\cite{gomez} & $\{,\}_i$ & $a_i$ & $b_i$ & $c_i$ \\ 
     \hline
   15 $(\ast)$ & $\{,\}_1$ & $-\omega$ & $0$ & $0$ \\  
   \hline 
   16  & $\{,\}_2$ & $-\lambda$ & $\cdot$ & $\cdot$ \\
    \hline
    \end{tabular}
      
    {\bf Table 7}. Classification of PL structures on $A_{3,7}$.
    
\end{center}
\item Remarks. The $\mu\to 0$ limit of the Lie algebra $A_{3,7}$ is just the Euclidean Lie algebra $A_{3,6}$. Therefore, the former can be thought of as a deformation of the latter (see the $A_{3,7}$ group element).
Note that the PL brackets are always quadratic in the group matrix entries, provided the constraint $C^2+S^2-\exp \left( -2 \alpha \arctan \left( \frac{S}C \right) \right)=0$ is imposed. As it has been stressed in~\cite{gomez}, when $\mu=1$ the non-coboundary Lie bialgebra $16$ is self-dual.

\end{enumerate}

\subsection{PL structures on the $SL(2,\mathds{R})$ group generated by  $A_{3,8}$}
\begin{enumerate}
\item Commutation relations
$$
[\We_1,\We_3]=-2 \We_2, \qquad [\We_1,\We_2]=\We_1, \qquad [\We_2,\We_3]=\We_3.
$$
\item Representation
$$
\varrho(\We_1)=\left( 
\begin{array}{cc}
0 & 0\\
1 & 0
\end{array}
\right),  \qquad 
\varrho(\We_2)=\frac12 \left( 
\begin{array}{cc}
1 & 0 \\
0 & -1 
\end{array}
\right), \qquad
\varrho(\We_3)=\left( 
\begin{array}{cc}
0 & 1 \\
0 & 0
\end{array}
\right).
$$
\item Matrix group element
$$
M=
\left( 
\begin{array}{cc}
e^{\frac{y}{2}} & x e^{\frac{y}{2}} \\
z e^{\frac{y}{2}} & x z e^{\frac{y}{2}} +e^{-\frac{y}{2}}
\end{array}
\right)=\left(
\begin{array}{cc}
Y & X\\
Z & W
\end{array} 
\right), \qquad Y>0, \ YW-XZ=1.
$$ 

\item Coproduct
\begin{eqnarray*}
\Delta(X)&=&X \otimes W+ Y \otimes X\\
\Delta(Y)&=&Y \otimes Y+ X \otimes Z,\\
\Delta(Z)&=&Z \otimes Y + W \otimes Z,\\
\Delta(W)&=&Z \otimes X + W \otimes W.
\end{eqnarray*}
\item Poisson--Lie bracket
\begin{eqnarray*}
\{ X,Y \}_1&=&-a_1 X^2+b_1 X Y+c_1 (1-Y^2), \\
\{ X,Z \}_1&=& -(a_1 X+c_1 Z) ( Y+W),\\
\{ X,W \}_1&=& -a_1 X^2-b_1 X W + c_1(1-W^2), \\
\{ Y,Z \}_1&=& a_1 (1-Y^2)-b_1 Y Z-c_1 Z^2, \\
\{ Y,W \}_1&=& 2 b_1(1-WY)+ (c_1 Z - a_1 X)(Y-W),\\
\{ Z,W \}_1&=& a_1 (W^2 -1)-b_1 ZW +c_1 Z^2.
\end{eqnarray*}

\item Casimir
$$
\mathcal{C}=\left\{
\begin{array}{l}
\displaystyle \dfrac{a_1(W-Y)-b_1 Z}{(a_1 X+c_1 Z)} \qquad (a_1\neq 0\ \mbox{or}\  c_1\neq 0)\\ \\
\displaystyle \dfrac{W-Y}{Z} \qquad (a_1=b_1=0) \\ \\
\displaystyle \dfrac{X}{Z} \qquad (a_1=c_1=0)
\end{array}
\right.
$$
\item Coboundary cases.
Since the Lie algebra $A_{3,8}$ is simple, the PL tensor is always a coboundary one, with the following $r-$matrix:
$$
 r=r^{12} \, \We_1 \wedge \We_2 +r^{13} \, \We_1 \wedge \We_3 + r^{23}  \, \We_2 \wedge \We_3.
$$
By computing the Sklyanin bracket we get the following identification for the parameters:
$$
a_1=-r^{12}, \qquad b_1=-2 r^{13}, \qquad c_1=r^{23}.
$$
\item Isomorphic to the algebra  $sl(2,\mathds{R})$ in~\cite{gomez} through the change of basis
$$
\We_1=\sqrt{2} \Ge_1 \qquad \We_2=-\Ge_0 \qquad  \We_3=\sqrt{2} \Ge_2.
$$ 
\item Correspondence with the classification of Lie bialgebras:
\begin{center}
    \begin{tabular}{ c | c | c | c | c |}
    Lie bialg.~in~\cite{gomez} & $\{,\}_i$ & $a_i$ & $b_i$ & $c_i$ \\ 
     \hline
    1 $(\ast)$ & $\{,\}_1$ & $0$ & $\lambda/2$ & $0$ \\   
    \hline
    2 $(\ast)$ & $\{,\}_1$ & $0$ & $0$ & $\sqrt{2} \lambda /4$  \\
    \hline
    3 $(\ast)$ & $\{,\}_1$ & $\sqrt{2} \lambda /4$ & $0$ & $0$  \\
    \hline
    \end{tabular}
      
    {\bf Table 8}. Classification of PL structures on $A_{3,8}$.
    
\end{center}
\item Remarks. The Lie bialgebra $1$ corresponds to the PL structure underlying the standard quantum $SL(2,\mathds{R})$ group, case $2$ corresponds to the standard quantum deformation of $SO(2,1)$ and the case 3 corresponds to the non-standard quantum  $SL(2,\mathds{R})$ group.
\end{enumerate}

\subsection{PL structures on the $SO(3)$ group generated by $A_{3,9}$}
\begin{enumerate}
\item Commutation relations
$$
[\We_1,\We_2]=\We_3, \qquad [\We_2,\We_3]=\We_1, \qquad [\We_3,\We_1]=\We_2
$$
\item Representation
$$
\varrho(\We_1)=\left( 
\begin{array}{ccc}
0 & 0 & 0 \\
0 & 0 & -1\\
0 & 1 & 0
\end{array}
\right),  \qquad 
\varrho(\We_2)=\left( 
\begin{array}{ccc}
0 & 0 & 1\\
0 & 0 & 0\\
-1 & 0 & 0
\end{array}
\right), \qquad
\varrho(\We_3)=\left( 
\begin{array}{ccc}
0 & -1 & 0 \\
1 & 0 & 0 \\
0 & 0 & 0
\end{array}
\right).
$$
\item Matrix group element
$$
M=
\left( 
\begin{array}{ccc}
C_x C_y & -S_x C_y & S_y\\
C_x S_y S_z + S_x C_z  &-S_x S_y S_z +C_x C_z & -C_y S_z\\
-C_x S_y C_z + S_x S_z & S_x S_y C_z + C_x S_z & C_y C_z 
\end{array}
\right).
$$
where we used the shorthand notation
$
C_\alpha=\cos(\alpha)$ and  $S_\alpha=\sin(\alpha).
$
\item Coproduct. It would be immediate to write it for the matrix entries of M, and in terms of the local group coordinates it can be formally written as
{\small
\begin{eqnarray*}
\Delta(x)&=& \pi-\arccos \left( \frac{-C_x C_y \otimes C_x C_y+ S_x C_y \otimes  C_x S_y S_z +S_x C_y \otimes S_x C_z +S_y \otimes C_x S_y C_z -S_y \otimes S_x S_z}{\sqrt{1-\sin(\Delta(y))}} \right),\\
\Delta(y)&=& \arcsin \left(C_x C_y  \otimes S_y +S_x C_y \otimes C_y S_z + S_y \otimes C_y C_z \right), \\
\Delta(z)&=& \pi-\arccos\left ( \frac{ C_x S_y C_z \otimes S_y  - S_x S_z \otimes  S_y+ S_x S_y C_z \otimes C_y S_z + C_x S_z \otimes C_y S_z - C_y C_z \otimes C_y C_z}
{\sqrt{1-\sin(\Delta(y))}} \right).
\end{eqnarray*}}
\normalsize

\item Poisson--Lie bracket. The most general Poisson bracket compatible with the above coproduct maps is given by
\begin{eqnarray*}
\{x,y\}_1 &=& \frac{a_1 \sin(y)+b_1 \sin(x) \cos(y)+c_1 \cos(x) \cos(y)-c_1}{\cos(y)},\\
\{x,z \}_1 &=& \frac{a_1 \sin(z)+b_1 \cos(x)-b_1 \cos(z)-c_1 \sin(x)}{\cos(y)},\\
\{ y, z \}_1&=&-\frac{a_1 \cos(z) \cos(y)+b_1 \sin(z) \cos(y)-a_1+c_1 \sin(y)}{\cos(y)}.
\end{eqnarray*}

\item Casimir. In the case $b_1=c_1=0$ (which is the only essential PL bracket, see below) it reads
$$
\mathcal{C}=\frac{x}{2}-\arctan \left[ \dfrac{\sin\left( \frac{y-z}{2} \right)}{\sin\left( \frac{y+z}{2} \right)} \right].
$$

\item Coboundary cases.
Again, the Lie algebra $A_{3,9}$ is simple, so its PL tensors are always coboundaries. The generic $r$-matrix is given by
\begin{equation*}
r=r^{12} \, \We_1 \wedge \We_2 + r^{13}  \, \We_1 \wedge \We_3 + r^{23}  \, \We_2 \wedge \We_3.
\end{equation*}
which means that
\begin{equation*}
a=r^{12}, \qquad b=r^{13}, \qquad c=r^{23}.
\end{equation*}

\item Isomorphic to the algebra  $so(3)$ in~\cite{gomez} through the change of basis
$$
\We_1= \Ge_1 \qquad \We_2=\Ge_2 \qquad  \We_3=\Ge_0.
$$ 
\item Correspondence with the classification of Lie bialgebras:
\begin{center}
    \begin{tabular}{ c | c | c | c | c |}
    Lie bialg.~in~\cite{gomez} & $\{,\}$ & $a$ & $b$ & $c$ \\ 
     \hline
    4 & $\{,\}_1$ & $\lambda$ & $0$ & $0$ \\   
    \hline
    \end{tabular}
      
    {\bf Table 9}. Classification of PL structures on $A_{3,9}$.
    
\end{center}
\item Remarks. The Lie bialgebra 4 is the one generated by the classical $r$-matrix $
r=a_1 \, \We_1 \wedge \We_2 $ (which is equivalent to the generic three-parametric $r$-matrix through the appropriate automorphism), whose Sklyanin bracket is given above and provides the semiclassical limit of the quantum $SO(3)$ group.

\end{enumerate}

\section{Conclusions} \label{conclusions}

In this paper we have constructed and classified all the possible PL structures for the nine real 3D Lie groups. For each of the PL brackets, we have given the explicit expressions for the coproduct map and for the corresponding Casimir function. Moreover, our results are fully consistent with the complete classification of Lie bialgebra structures given in~\cite{gomez}, that we have used in order to identify all the unequivalent  PL structures under generic group automorphisms. 

For each Lie group, the PL structures are obtained by solving the cocycle condition through direct computation and by assuming initially a quadratic dependence of the PL bracket in terms of the group matrix entries (this assumption has to be relaxed in only two cases). The solutions so obtained are grouped into multiparametric families of Poisson brackets, that in many cases provide non-coboundary PL structures that had not been constructed so far. In particular, the PL structures for the groups corresponding to the solvable Lie algebras $A_{3,2}$, $A_{3,4}$ (the (1+1) Poincar\'e algebra), $A_{3,5}$ and $A_{3,7}$ are -to the best of our knowledge- presented here for the first time.

The approach here presented can be straightforwardly implemented in order to obtain and classify PL structures of non-semisimple Lie groups in higher dimensions, for which known results are scarce and essentialy deal with coboundary structures~\cite{anna,BCHgalilei,BHosc,BHPSchrod, kuperb}. In particular, it is known that all the Lie bialgebra structures 
of groups built as semidirect products between (2+1) and (3+1) space-time rotations (with arbitrary signature) and 
translations, are coboundaries~\cite{zakr}, although only the (2+1)~\cite{stachura} and the (3+1) Poincar\'e
classification~\cite{zakr,woro,zakrcmp} were completed. However, a bunch of non-coboundary Lie bialgebras arise in the classification of the massless (3+1) Galilei PL groups~\cite{brihaye}
and similar results can be expected for the classification of the PL structures on the 4D and 5D real Lie groups whose Lie algebras are classified in~\cite{pavel}.

Consequently, the results here presented can be useful in two different directions. On one hand, they provide a complete and closed chart of the semiclassical counterparts of all possible 3D real quantum groups, that in many non-semisimple (and non-coboundary) cases are yet unexplored. On the other hand, as in the case of Lotka-Volterra equations~\cite{LV}, it could happen that other relevant dynamical systems whose Hamiltonian structure is provided by quadratic Poisson algebras could find a group theoretical interpretation as Poisson-Lie structures on certain (possibly non-simple) Lie groups. Work on these two lines is in progress.

\section*{Acknowledgements}

This work was partially supported by the Spanish MICINN   under grant   MTM2010-18556
and by INFN--MICINN (grant AIC-D-2011-0711).

\end{document}